\documentclass[aps,preprint,prd,tightenlines]{revtex4}
\usepackage{graphics}
\begin{document}
\title{Large {\protect\boldmath{$N$}} planar 
or bare vertex approximation and 
critical behavior of a SU({\protect\boldmath{$N$}}) invariant 
four-fermion model in 2+1 dimensions}
\author{Manuel Reenders}
\affiliation{
Department of Polymer Chemistry and Materials Science Center,\\
University of Groningen, Nijenborgh 4,
9747 AG Groningen, The Netherlands}
\date{\today}
\begin{abstract}
A four-fermion model in 2+1 dimensions describing $N$ Dirac fermions
interacting via SU($N$) invariant $N^2-1$ four-fermion interactions is
solved in the leading order of the $1/N$ expansion. The $1/N$
expansion corresponds to 't~Hoofts topological $1/N$ expansion in
which planar Feynman diagrams prevail.  For the symmetric phase of
this model, it is argued that the planar expansion corresponds to the
ladder approximation.  A truncated set of Schwinger-Dyson equations
for the fermion propagator and composite boson propagator representing
the relevant planar diagrams is solved analytically. The critical
four-fermion coupling and various critical exponents are determined as
functions of $N$.  The universality class of this model turns out to
be quite distinct from the Gross-Neveu model in the large $N$ limit.
\end{abstract}
\maketitle
%
%
%
\section{Introduction}\label{sec_intro}
Already in the early seventies K.G.~Wilson \cite{wi73} argued that
four-fermion theories in dimensions $2<d<4$ are nontrivially
renormalizable at least in leading order of the $1/N$ expansion, with
$N$ the number of fermion flavors.  The model studied by Wilson is
nowadays referred to as the $d$ dimensional generalization of
Gross-Neveu model \cite{grne74}.  In the Gross-Neveu model with $N$
fermion flavors there is one composite or auxiliary field
$\sigma\propto -\sum_{\alpha=1}^N\bar\psi_\alpha\psi_\alpha$
describing the scalar bound states; the number of fermions is taken to
be much larger than the number of light bound states.  In the $1/N$
expansion for such models, Feynman diagrams with fermion loops
dominate over other types of diagrams.

The Gross-Neveu model in 2+1 dimensions is well-known to exhibit
dynamically symmetry breaking whenever the four-fermion coupling
exceeds some critical value.  About a decade ago renewed interest in
the Gross-Neveu model led to a number of papers establishing the
renormalizability of the Gross-Neveu model and related models in the
$1/N$ expansion due to the presence of a finite ultraviolet stable
fixed point and conformal invariance at this critical point
\cite{rowapa89arowa89,sewij89,kiya90,hakoko91,chmase93}.  Probably the
most extensive study of the Gross-Neveu model, was performed by Hands,
Koci\'c, and Kogut \cite{hakoko93}.  In their paper analytical results
in next-to-leading order in $1/N$ were verified in numerical lattice
simulations. The 1+1 dimensional four-fermion models cannot exhibit
dynamical symmetry breaking, due to the Coleman-Mermin-Wagner theorem,
whereas $3+1$ dimensional models usually suffer from the triviality
problem \cite{mirbook,re9900}.

In this paper a SU($N$)$\times$ U(1) invariant four-fermion model,
closely related to the Gross-Neveu model, is studied in 2+1
dimensions.  In this model the four-fermionic potential contains
$N^2-1$ terms, {\em i.e.}, the four-fermion interactions are in the
adjoint representation of the SU($N$) symmetry. Thus the model
presented here describes the interaction of $N$ Dirac fermions with
$N^2-1$ scalar composite states.  Consequently the large $N$ treatment
will be quite different from the Gross-Neveu model.  Instead of the
large $N$ expansion \`a la Wilson, 't Hoofts topological $1/N$
expansion \cite{tho74} will be adopted.  We assume that the $1/N$
expansion of 't Hooft is particularly useful for theories with a
global U($N$) symmetry containing fields with two U($N$) or SU($N$)
indices, {\em e.g.}, see Refs.~\cite{mirbook,re9900,tho74,kl92,fo00afo00b}.
In such an expansion planar diagrams dominate over diagrams with
topologies other than planar.

The leading large $N$ critical behavior of the proposed model is
studied, with the main focus on the determination of the scaling
behavior of the fermion wave function and of the $N^2-1$ scalar or
$\sigma$ boson propagators.  Once the anomalous dimensions of the
fermion propagator and the $\sigma$ boson propagators are determined,
the gap equation for the order parameter will be studied.  The
dynamical breaking of the SU($N$) symmetry to U(1) gives rise to the
Goldstone realization with $N^2-2$ Nambu-Goldstone bosons and one
massive scalar in the broken phase.

The main motivation for studying this particular three dimensional
four-fermionic model is it possible relationship with low temperature
behavior of the 2+1 dimensional Hubbard-Heisenberg and $t-J$ models
near the antiferromagnetic wave vector
\cite{fo00afo00b,afma88maaf89,albomasij00,re02}.

The setup of the paper is the following.  The model Lagrangian is
introduced in the next section.  We present a truncation scheme for
the Schwinger-Dyson (SD) equations which generates all planar and thus
leading $1/N$ diagrams in Sec.~\ref{sec_largeN}.  This closed set of
SD equations is subsequently solved analytically in
Sec.~\ref{sec_scalbeh}.  In Sec.~\ref{sec_gapeq} the dynamical
symmetry breaking is analyzed and the critical exponents,
hyperscaling, and universality are discussed.  Next-to-leading order
$1/N$ corrections are discussed in Sec.~\ref{sec_dis}.  The
conclusions are presented in Sec.~\ref{sec_concl}.  In the
Appendix~\ref{ap_1overN}, the applicability of the topological $1/N$
expansion is motivated for this 2+1 dimensional model.
\section{The Model}\label{sec_model}
We will consider the following 2+1 dimensional four-fermion model
given by the Lagrangian
\begin{equation}
{\cal L}=\bar\psi_\alpha i\hat\partial\psi_\alpha
+\frac{G}{2}\sum_{A=1}^{N^2-1}(\bar\psi_\alpha
\tau^A_{\alpha\beta}\psi_\beta)^2,
\label{sunffm}
\end{equation}
where the index $\alpha=1,\dots,N$ labels the $N$ fermion flavors.
The fermions are described by four-component Dirac spinors.  The
Lagrangian Eq.~(\ref{sunffm}) is globally invariant under
SU($N$)$\times$ U(1), see also Ref.~\cite{kl92}. By the way, in
Refs.~\cite{mirbook,re9900} a U($N$)$\times$ U($N$) invariant version
of Eq.~(\ref{sunffm}) is described in $3+1$ dimensions.  The Hermitian
generators, $\tau^A$, of the SU($N$) symmetry satisfy
\begin{eqnarray}
&&{\rm Tr}[\tau^A]=0,\quad {\rm Tr}[\tau^A\tau^B]=\delta^{AB},\quad
\sum_{A=1}^{N^2-1}\tau^A_{\alpha\beta}\tau^A_{\gamma\delta}
=\delta_{\alpha \delta}\delta_{\gamma\beta}
-\frac{1}{N}\delta_{\alpha\beta}\delta_{\gamma\delta}.
\end{eqnarray}
After a Hubbard-Stratonovich transformation, the Lagrangian
(\ref{sunffm}) reads
\begin{eqnarray}
{\cal L}&=&
\bar\psi_\alpha i\hat \partial\psi_\alpha
-\sum_A\bar\psi_\alpha\tau^A_{\alpha\beta}\psi_\beta\sigma_A
-\frac{1}{2G}\sum_A\sigma_A^2,\label{HSsunffm}
\end{eqnarray}
with the real auxiliary bosonic fields $\sigma^A=-G\bar\psi\tau^A\psi$
conveniently describing the composite fermion--antifermionic degrees
of freedom. Another equivalent representation with $\sigma_{ij}=\sum_A
\sigma^A \tau^A_{ij}$ can be adopted \cite{kl92}, see also the
Appendix~\ref{ap_1overN}.  Along the lines of Ref.~\cite{re9900},
$N^2-1$ $\sigma$ boson propagators $\Delta_\sigma^A(p)$ and Yukawa
vertices $\Gamma^A_\sigma(p+q,p)$ (``fully amputated'') are
introduced. The bare Yukawa vertices are $\gamma^A_\sigma=\tau^A {\bf
1}$.  Since in the symmetric phase the $\Delta^A_\sigma$ propagators
are identical, the index $A$ is often discarded.

The dynamical symmetry breaking SU($N$)$\,\rightarrow\,$U(1) is
described by the order parameter $\langle \sigma_3 \rangle\simeq
-G\langle\bar\psi\tau^3\psi\rangle$, which is a particular choice for
the symmetry breaking term. In principle we have the freedom to choose
from $N^2-1$ order parameters $\bar\psi\tau^A\psi$.  In order to
``force'' the symmetry breaking in the $\tau^3$ direction a bare mass
term $-h\bar\psi\tau^3\psi$ is added to the Lagrangian in
Eq.~(\ref{HSsunffm}). In the end the limit $h\rightarrow 0$ can be
taken.

In the fermion propagator the symmetry breaking is given by the mass
function $M$ which is introduced as follows
\begin{eqnarray}
S_{\alpha\beta}(p)=\frac{A(p)\hat p\delta_{\alpha\beta} +\sqrt{N}
M\tau^3_{\alpha\beta}}{A^2(p)p^2-M^2}.\label{fermprop}
\end{eqnarray}
The function $A(p)$ is referred to as the fermion wave function.  The
fermion dynamical mass $m_{dyn}$ is defined as the mass pole in the
fermion propagator; it is related to $M$ via
$A^2(m_{dyn})m^2_{dyn}-M^2$.  Both the fermion dynamical mass
$m_{dyn}$ or the mass pole $m_\sigma$ in the $\sigma$ boson propagator
can be considered as the appropriate inverse correlation length
characterizing the infrared mass scale of this model.
\section{Large {\protect{\boldmath$N$}} planar 
and bare vertex approximation}
\label{sec_largeN}
The present model describes $N$ species of fermions interacting via
$N^2-1$ four-fermionic interactions in the so-called adjoint
representation of the SU($N$) flavor symmetry.  Therefore, when $N$ is
large, we assume the applicability of 't Hoofts topological $1/N$
expansion \cite{tho74}, see also Ref.~\cite{re9900}.  Moreover, since
the U($N$) symmetry is global, we assume that the corresponding
topological $1/N$ expansion is independent of the space-time dimension
$d$.  The applicability of 't Hoofts $1/N$ expansion to the present
model is motivated in the Appendix~\ref{ap_1overN}.  The topological
$1/N$ expansion states that the Feynman diagrams can be classified in
terms of surfaces with a specific topology.  The so-called planar
Feynman diagrams, with the fermions at the boundary, describe the
leading contributions to the Green functions.  Diagrams with other
than planar topological structures are multiplied by factors of at
least $1/N$, and in the limit of large $N$, their contribution can be
neglected with respect to planar graphs.  For instance diagrams with
one handle on their representing surface are suppressed by a factor
$1/N^2$ relatively to their planar counter parts, whereas diagrams
with a hole are suppressed by a factor $1/N$.

\begin{figure}
\resizebox*{0.8\columnwidth}{!}{\includegraphics{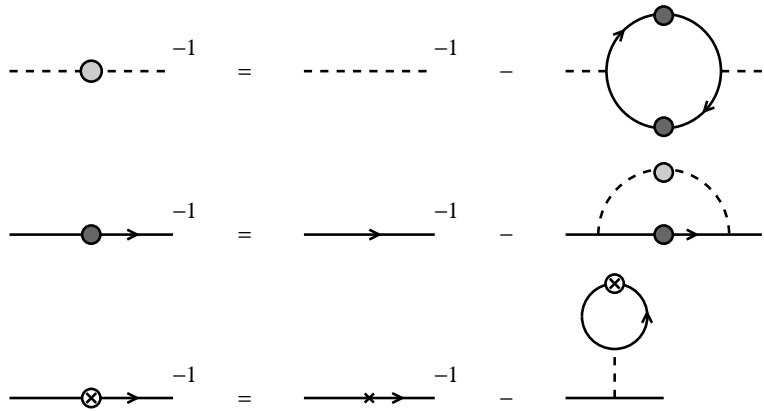}}
\caption{The planar approximation; the dashed lines are the real
$\sigma$ boson propagators, the black lines are fermion propagators.
The blobs correspond to the fully dressed propagators.  The dark blob
in the fermion propagator corresponds to a wave function part being
proportional to $\hat p$, whereas the white blob with a cross
corresponds to a mass insertion proportional to the $\tau^3$
generator.  }
\label{fig_trunc}
\end{figure}

\begin{figure}
\resizebox*{0.45\columnwidth}{!}{\includegraphics{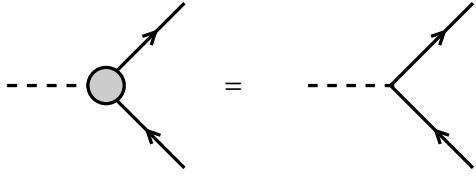}}
\caption{The bare vertex approximation.}
\label{fig_barevertex}
\end{figure}

In this paper we are mainly interested in the SD equations for the
$\sigma$ boson propagator $\Delta_\sigma$ and the fermion wave
function $A$.  In the Appendix~\ref{ap_1overN}, it is shown that the
leading large $N$ behavior for $\Delta_\sigma$ is given by those
diagrams which do not contain Yukawa vertex ($\Gamma_\sigma$)
corrections, see the first two truncated SD equations in
Fig.~\ref{fig_trunc}.
In other words, only self energy corrections ($\Sigma$) and $\sigma$
boson vacuum polarization corrections ($\Pi_\sigma$) should be
considered.  One can straightforwardly verify that the bare vertex
approximation (see Fig.~\ref{fig_barevertex}),
$\Gamma^A_\sigma(p+q,p)\approx \tau^A {\bf 1}$, generates only planar
and thus leading large $N$ graphs for $\Delta_\sigma$ and $A$.  Here
we mention that the propagators of the real scalar fields $\sigma^A$
are represented by dashed lines, whereas the propagators of the
Hermitian fields $\sigma_{\alpha\beta}$ of the
Appendix~\ref{ap_1overN} are represented by double lines.

For the symmetry breaking fermion mass $M$ only the tad-pole graph
contributes in the leading large $N$ limit, see Fig.~\ref{fig_trunc}.
At a first glance this might seem inconsistent, since parts of the
fermion propagator are now described by different truncations.
However, the self-consistency of the truncated set of SDEs given in
Fig.~\ref{fig_trunc} can be verified using the SU($N$) Ward
identities, {\em e.g.}, see Ref.~\cite{re9900}.  These Ward identities
relate the $N^2-1$ Yukawa vertices $\Gamma^A_\sigma(p,p)$ at zero
$\sigma$ boson momentum transfer to the fermion dynamical mass $M(p)$
in the following way:
\begin{eqnarray}
\Gamma^A_\sigma(p,p)&=&\tau^A \frac{M(p)}{\langle\sigma_3\rangle},
\quad A=1,\dots N^2,\quad A\not=3,\\
\Gamma^3_\sigma(p,p)&=&\tau^3 
\frac{\partial M(p)}{\partial \langle \sigma_3\rangle},
\end{eqnarray}
when the symmetry breaking is in the $\tau^3$ direction.  Thus if we
apply the bare vertex approximation $\Gamma^A_\sigma=\tau^A$
(Fig.~\ref{fig_barevertex}), we get the equation $M=\langle
\sigma_3\rangle$.  This equation is indeed represented by the third SD
equation in Fig.~\ref{fig_trunc}.

Summarizing; the leading large $N$ behavior can be described
consistently in terms of the truncated set of SDEs given in
Fig.~\ref{fig_trunc}.
\section{Scaling behavior at criticality}\label{sec_scalbeh}
Having introduced the leading large $N$ planar or bare vertex
approximation for this model, we proceed by studying the truncated set
given in Fig.~\ref{fig_trunc} in the symmetric phase first.  These
equations are in Euclidean formulation
\begin{eqnarray}
\Delta_\sigma^{-1}(p)&=&-\frac{1}{G}+\Pi_\sigma(p),\label{Delsprop}\\
\Pi_\sigma(p)&=&\int_E\frac{d^3k}{(2\pi)^3}\,\frac{4(k^2+k\cdot
p)}{(k+p)^2 k^2 A(k+p) A(k)},
\label{vacpoleq}
\end{eqnarray}
and 
\begin{eqnarray}
A(p)&=&1-N\Sigma_A(p),\label{Aeq}\\
\Sigma_A(p)&=&
\int_E \frac{d^3k}{(2\pi)^3}\,
\frac{p\cdot k}{p^2 k^2 A(k)}\Delta_\sigma(k-p).
\label{sigAdef}
\end{eqnarray}
For the dimensionless coupling constant $g\equiv 2G\Lambda/\pi^2$ 
there exists an ultraviolet stable fixed point $g_c$ above which the SU($N$) 
is dynamically broken. Exactly at the critical coupling $g=g_c$ 
the infrared mass scale vanishes and 
the above propagators will be given in terms of pure power-laws.
Therefore, we first consider the two coupled SD equations 
at the critical point ($g=g_c$), where we can safely
assume a power-law form for the fermion wave function
\begin{equation}
A(p)=(p^2/\Lambda^2)^{-\zeta/2}, \label{wavefieansatz}
\end{equation}
where $\Lambda$ is the ultraviolet cutoff, and 
where both the critical coupling $g_c$ and
the exponent $\zeta$ are to be determined from the above SD equations
and will depend on $N$ via Eq.~(\ref{Aeq}). 
Using the above ansatz for the fermion wave function, we subsequently
solve the SD for the $\sigma$ boson propagator.
For the model to be nontrivially renormalizable, 
the $\sigma$ boson propagator should also exhibit a pure power-law form
at the critical coupling;
\begin{eqnarray}
\Delta_\sigma(p)\propto \left(\frac{\Lambda}{p}\right)^{2-\eta},
\label{Delsansatz}
\end{eqnarray}
where $\eta$ is the anomalous dimension of the $\sigma$ boson propagator.
Therefore, in what follows, 
we solve the SD equations (\ref{Delsprop})--(\ref{sigAdef}) 
assuming the power-law forms (\ref{wavefieansatz}) and (\ref{Delsansatz})
in order to obtain the exponents $\zeta$ and $\eta$ and 
the critical coupling $g_c$ as function of $N$. 

Inserting Eq.~(\ref{wavefieansatz}) in Eq.~(\ref{vacpoleq}) we obtain
the integral
\begin{eqnarray}
\Pi_\sigma(p)=\frac{2\Lambda^{-2\zeta}}{\pi^2}\int\limits_0^\Lambda dk\,
\int\frac{d\Omega}{4\pi}\,
\frac{(k^2+k\cdot p)}{k^{-\zeta}(k^2+p^2+2 k\cdot p)^{1-\zeta/2}}.
\end{eqnarray}
The angular integral is straightforward
\begin{eqnarray}
\Pi_\sigma(p)&=& \frac{2 \Lambda^{-2\zeta}}{\pi^2} \int\limits^{\Lambda}_0
dk\,\frac{k^\zeta}{2\zeta (2+\zeta)kp} \biggr[(k^2+\zeta k^2+\zeta
kp-p^2)(k+p)^\zeta\nonumber\\ &-& (k^2+\zeta k^2-\zeta
kp-p^2)(|k-p|)^\zeta \biggr],
\end{eqnarray}
where $k=\sqrt{k^2}$ and $p=\sqrt{p^2}$.
We split this integral into three parts;
\begin{eqnarray}
\Pi_\sigma(p)= \left(\frac{2\Lambda}{\pi^2}\right)
\left(\frac{p}{\Lambda}\right)^{1+2\zeta}\left[{\cal J}_1(\zeta,x)
+{\cal J}_2(\zeta,x)+{\cal J}_3(\zeta,x)\right],
\label{iintdef}
\end{eqnarray}
with $x=p/\Lambda$ and
where
\begin{eqnarray}
{\cal J}_1(\zeta,x)&\equiv&\frac{(1+\zeta)}{2\zeta(2+\zeta)}
\left[\int\limits_0^1dt\,t^{\zeta+1} f_1(t)
+\int\limits_x^1
dt\,t^{-2\zeta-3}f_1(t)\right],
\\ {\cal J}_2(\zeta,x)&\equiv&\frac{1}{2(2+\zeta)}
\left[\int\limits_0^1dt\,t^{\zeta}f_2(t) +\int\limits_x^1
dt\,t^{-2\zeta-2}f_2(t)\right],\\
{\cal J}_3(\zeta,x)&\equiv&-\frac{1}{2\zeta(2+\zeta)}
\left[\int\limits_0^1dt\,t^{\zeta-1}f_1(t) +\int\limits_x^1
dt\,t^{-2\zeta-1}f_1(t)\right],
\end{eqnarray}
with
\begin{eqnarray}
f_1(t)\equiv (1+t)^\zeta-(1-t)^\zeta,\qquad
f_2(t)\equiv (1+t)^\zeta+(1-t)^\zeta.
\end{eqnarray}
The integrals can be performed straightforwardly using the 
Table of Integrals \cite{Gradstein}
and the
${\cal J}$ function can be expressed in terms of the $\Gamma$ function and
the hypergeometric function $_2F_1=F$. 
In this way, the function ${\cal J}_1$ can be expressed as
(with $x=p/\Lambda$)
\begin{eqnarray}
{\cal J}_1(\zeta,x)&=&\frac{(1+\zeta)}{2\zeta(2+\zeta)} \Biggr\{
\nonumber\\&&
\frac{x^{-2\zeta-2}}{(2+2\zeta)}
\left[F(-2-2\zeta,-\zeta;-1-2\zeta;-x)-F(-2-2\zeta,-\zeta;-1-2\zeta;x)
\right]\nonumber\\
&+&\frac{1}{(2+\zeta)}F(2+\zeta,-\zeta;3+\zeta;-1)
-\frac{1}{(2+2\zeta)}F(-2-2\zeta,-\zeta;-1-2\zeta;-1)
\nonumber\\
&-&\frac{\Gamma(2+\zeta)\Gamma(1+\zeta)}{\Gamma(3+2\zeta)}
-\frac{\Gamma(-2-2\zeta)\Gamma(1+\zeta)}{\Gamma(-1-\zeta)}
\Biggr\}.
\end{eqnarray}
Subsequently ${\cal J}_2(\zeta,x)$ and ${\cal J}_3(\zeta,x)$ 
are computed in a similar manner, giving
\begin{eqnarray}
{\cal J}_2(\zeta,x)&=&\frac{1}{2(2+\zeta)}
\Biggr\{
\frac{x^{-2\zeta-1}}{(1+2\zeta)}
\left[F(-1-2\zeta,-\zeta;-2\zeta;-x)+F(-1-2\zeta,-\zeta;-2\zeta;x)
\right]\nonumber\\
&+&\frac{1}{(1+\zeta)}F(1+\zeta,-\zeta;2+\zeta;-1)
-\frac{1}{(1+2\zeta)}F(-1-2\zeta,-\zeta;-2\zeta;-1)
\nonumber\\
&+&\frac{\Gamma(1+\zeta)\Gamma(1+\zeta)}{\Gamma(2+2\zeta)}
+\frac{\Gamma(-1-2\zeta)\Gamma(1+\zeta)}{\Gamma(-\zeta)}
\Biggr\},
\end{eqnarray}
and
\begin{eqnarray}
{\cal J}_3(\zeta,x)&=&-\frac{1}{2\zeta(2+\zeta)}
\Biggr\{
\frac{x^{-2\zeta}}{2\zeta}
\left[F(-2\zeta,-\zeta;1-2\zeta;-x)-F(-2\zeta,-\zeta;1-2\zeta;x)
\right]\nonumber\\
&+&\frac{1}{\zeta}F(\zeta,-\zeta;1+\zeta;-1)
-\frac{1}{2\zeta}F(-2\zeta,-\zeta;1-2\zeta;-1)
\nonumber\\
&-&\frac{\Gamma(\zeta)\Gamma(1+\zeta)}{\Gamma(1+2\zeta)}
-\frac{\Gamma(-2\zeta)\Gamma(1+\zeta)}{\Gamma(1-\zeta)}
\Biggr\}.
\end{eqnarray}
The condition for the existence of these integrals is 
that $\zeta>0$.
Assuming that $0<\zeta <1/2$, we can determine 
the asymptotic behavior or infrared limit of the above ${\cal J}$ functions, 
by taking the $x\rightarrow 0$ limit ($p\ll \Lambda$).
We write
\begin{eqnarray}
{\cal J}_i(\zeta,x)&\simeq& {\cal A}_i(\zeta) x^{-1-2\zeta}-{\cal B}_i(\zeta)
+{\cal C}_i(\zeta)x^{1-2\zeta}+\dots,
\qquad i=1,\,2,\,3,
\end{eqnarray}
or
\begin{eqnarray}
{\cal J}(\zeta,x)&\simeq& {\cal A}(\zeta) x^{-1-2\zeta}-{\cal B}(\zeta)+{\cal C}(\zeta)x^{1-2\zeta}+\dots,
\label{Jexpr1}
\end{eqnarray}
with ${\cal A}(\zeta)=\sum_{i=1}^3 {\cal A}_i(\zeta)$, ${\cal B}=\sum_{i=1}^3 {\cal B}_i$,
${\cal C}=\sum_{i=1}^3 {\cal C}_i$,
and where
\begin{eqnarray}
{\cal A}_1=\frac{1}{2(2+\zeta)}+\frac{1}{2(2+\zeta)(1+2\zeta)},\qquad
{\cal A}_2=\frac{1}{(2+\zeta)(1+2\zeta)},\qquad {\cal A}_3=0,
\end{eqnarray}
and
\begin{eqnarray}
{\cal B}_1(\zeta)&=&
-\frac{(1+\zeta)}{2\zeta(2+\zeta)} \Biggr\{
\frac{1}{(2+\zeta)}F(2+\zeta,-\zeta;3+\zeta;-1)\nonumber\\
&-&\frac{1}{(2+2\zeta)}F(-2-2\zeta,-\zeta;-1-2\zeta;-1)
-\frac{\Gamma(2+\zeta)\Gamma(1+\zeta)}{\Gamma(3+2\zeta)}
\nonumber\\
&-&\frac{\Gamma(-2-2\zeta)\Gamma(1+\zeta)}{\Gamma(-1-\zeta)}
\Biggr\}
,\\
{\cal B}_2(\zeta)&=&-\frac{1}{2(2+\zeta)}
\Biggr\{\frac{1}{(1+\zeta)}F(1+\zeta,-\zeta;2+\zeta;-1)
-\frac{1}{(1+2\zeta)}F(-1-2\zeta,-\zeta;-2\zeta;-1)
\nonumber\\
&+&\frac{\Gamma(1+\zeta)\Gamma(1+\zeta)}{\Gamma(2+2\zeta)}
+\frac{\Gamma(-1-2\zeta)\Gamma(1+\zeta)}{\Gamma(-\zeta)}
\Biggr\},
\end{eqnarray}
and
\begin{eqnarray}
{\cal B}_3(\zeta)&=&\frac{1}{2\zeta(2+\zeta)} \Biggr\{
\frac{1}{\zeta}F(\zeta,-\zeta;1+\zeta;-1)
-\frac{1}{2\zeta}F(-2\zeta,-\zeta;1-2\zeta;-1)
\nonumber\\
&-&\frac{\Gamma(\zeta)\Gamma(1+\zeta)}{\Gamma(1+2\zeta)}
-\frac{\Gamma(-2\zeta)\Gamma(1+\zeta)}{\Gamma(1-\zeta)}
\Biggr\}.
\end{eqnarray}
For $0<\zeta<1/2$, only the leading scaling terms or power-laws
$x^{-1-2\zeta}$ and $x^0$ of Eq.~(\ref{Jexpr1})
are relevant for our purposes, therefore the
functions ${\cal C}_i$ are not displayed.  
Then the vacuum polarization is
\begin{eqnarray}
\Pi_\sigma(p)\simeq \frac{2\Lambda}{\pi^2}\left[\frac{1}{1+2\zeta}-{\cal B}(\zeta) 
\left(\frac{p}{\Lambda}\right)^{1+2\zeta}\right],\label{Pisigexpr}
\end{eqnarray}
where ${\cal B}={\cal B}_1+{\cal B}_2+{\cal B}_3$.  
The critical coupling $g_c$ is given by 
the condition $\Delta_\sigma^{-1}(0)=0$ in Eq.~(\ref{Delsprop}), 
thus
\begin{equation}
\frac{1}{g_c}={\cal A}_1+{\cal A}_2+{\cal A}_3=\frac{1}{1+2\zeta}.\label{newcritg}
\end{equation}
With the expression (\ref{Pisigexpr}), the boson propagator
(Eq.~(\ref{Delsprop})) takes the pure power-law 
form (\ref{Delsansatz}) at the critical
coupling $g=g_c=1+2\zeta$:
\begin{eqnarray}
\Delta_\sigma(p)\simeq-\frac{\pi^2}{2\Lambda {\cal B}(\zeta)}
\left(\frac{\Lambda}{p}\right)^{1+2\zeta}, \label{bospropcrit}
\end{eqnarray}
consequently, the anomalous dimension $\eta=1-2\zeta$.

It can be shown that ${\cal B}_3(\zeta\rightarrow 0)\approx \pi^2/8$, and
that ${\cal B}_1$ and ${\cal B}_2$ approach zero when $\zeta\rightarrow 0$,
thus ${\cal B}(0)=\pi^2/8$.  
The case $\zeta=0$ ({\em i.e.} $A(p)=1$) 
in the above equation simply corresponds to the one-loop
or leading $1/N$ correction, $\Pi_\sigma(p)\simeq
(2\Lambda/\pi^2)(1-{\cal B}(0)p/\Lambda)$, see Ref.~\cite{hakoko91}.

At $\zeta=1/2$ it can be shown that ${\cal B}$ diverges, but so that 
\begin{eqnarray}
{\cal B}(\zeta\to 1/2)={\cal C}(\zeta\to 1/2),
\qquad {\cal B}(\zeta\to 1/2)\approx \frac{3}{8 (1-2\zeta)},
\label{Bnearzhalf}
\end{eqnarray}
and ${\cal J}(\zeta,x)$ reduces to
\begin{eqnarray}
{\cal J}(\zeta,x)\simeq \frac{1}{2 x^2}+\frac{3}{8}\ln x
\end{eqnarray}
at $\zeta=1/2$.
Consequently, at $\zeta=1/2$, 
the vacuum polarization $\Pi_\sigma$ of Eq.~(\ref{vacpoleq})
reads 
\begin{eqnarray}
\Pi_\sigma(p)\simeq \frac{2\Lambda}{\pi^2}\left[\frac{1}{2}
+\frac{3}{8}\frac{p^2}{\Lambda^2}
\ln\left(\frac{p}{\Lambda}\right)
\right].\label{Pilog}
\end{eqnarray}
This means there is no pure power-law form at $\zeta=1/2$ and consequently
the model becomes non-renormalizable in this case.
This situation resembles the situation in $3+1$ dimensional
four-fermion models such as the gauged Nambu--Jona-Lasinio model,
when the gauge coupling approaches zero, 
{\em e.g.}, see Ref.~\cite{aptewij91gure98}.

With the asymptotic form for the boson propagator,
Eq.~(\ref{bospropcrit}) and the Ansatz (\ref{wavefieansatz}) for the
wave function $A$, we can compute the Euclidean self-energy
contribution Eq.~(\ref{sigAdef}) and verify the self-consistency of 
Eq.~(\ref{wavefieansatz}).  Thus
\begin{eqnarray}
\Sigma_A(p)=-
\frac{\Lambda^{\zeta}}{4 {\cal B}(\zeta)}
\int\limits_0^{\Lambda}dk\,
\int\frac{d\Omega}{4\pi}\,\frac{k\cdot p}{p^2 k^{-\zeta} 
(k^2+p^2-2k\cdot p)^{1/2+\zeta}}.
\end{eqnarray}
Again the angular integral can be performed straightforwardly;
\begin{eqnarray}
\Sigma_A(p)&=&-
\frac{\Lambda^{\zeta}}{4 {\cal B}(\zeta) 2(3-2\zeta)(1-2\zeta)}
\int\limits_0^{\Lambda}dk\,
\frac{k^{\zeta-1}}{p^3}
\biggr\{
\left[k^2-(1-2\zeta)kp+p^2\right](k+p)^{1-2\zeta}\nonumber\\
&-&
\left[k^2+(1-2\zeta)kp+p^2\right](|k-p|)^{1-2\zeta}
\biggr\}.
\end{eqnarray}
The integral is expressed as follows:
\begin{eqnarray}
\Sigma_A(p)=-\frac{{\cal F}(\zeta,p/\Lambda)}{8 {\cal B}(\zeta) (3-2\zeta)(1-2\zeta)}
\left(\frac{p}{\Lambda}\right)^{-\zeta},
\end{eqnarray}
where
\begin{eqnarray}
{\cal F}(\zeta,p/\Lambda)
&\equiv&\int\limits_0^p\frac{dk}{p}\,\frac{k^{\zeta-1}}{p^{\zeta-1}}
\Biggr\{
\left[\frac{k^2}{p^2}-(1-2\zeta)\frac{k}{p}+1\right]
\left(1+\frac{k}{p}\right)^{1-2\zeta}\nonumber\\
&-&
\left[\frac{k^2}{p^2}+(1-2\zeta)\frac{k}{p}+1\right]
\left(1-\frac{k}{p}\right)^{1-2\zeta}\Biggr\}\nonumber\\
&+&
\int\limits_p^\Lambda\frac{dk}{k}\,\frac{k^{-\zeta+3}}{p^{-\zeta+3}}
\Biggr\{ \left[1-(1-2\zeta)\frac{p}{k}+\frac{p^2}{k^2}\right]
\left(1+\frac{p}{k}\right)^{1-2\zeta}\nonumber\\ &-&
\left[1+(1-2\zeta)\frac{p}{k}+\frac{p^2}{k^2}\right]
\left(1-\frac{p}{k}\right)^{1-2\zeta}\Biggr\}.
\end{eqnarray}
The function ${\cal F}$ is written as
\begin{eqnarray}
{\cal F}(\zeta,x)=\int\limits_0^1 dt\,t^{\zeta-1} \kappa(t)
+\int\limits_x^1 dt\, t^{\zeta-4} \kappa(t),\label{calFdef}
\end{eqnarray}
where
\begin{eqnarray}
\kappa(t)\equiv \left[t^2-(1-2\zeta)t+1\right]\left(1+t\right)^{1-2\zeta}
-\left[t^2+(1-2\zeta)t+1\right]\left(1-t\right)^{1-2\zeta}.\label{kapdef}
\end{eqnarray}
Performing the integral, and
determining the leading small $x=p/\Lambda$ behavior of the
function ${\cal F}$, we find for ${\cal F}$ 
\begin{eqnarray}
{\cal F}&=&\frac{\Gamma(\zeta)\Gamma(2-2\zeta)}{\Gamma(2-\zeta)}
\frac{(1+\zeta)(2\zeta-3)}{(3-\zeta)}
+\frac{1}{\zeta}F\left(\zeta,-1+2\zeta;1+\zeta;-1\right)\nonumber\\
&-&\frac{(1-2\zeta)}{1+\zeta}
F\left(1+\zeta,-1+2\zeta;2+\zeta;-1\right)
+\frac{1}{2+\zeta} F\left(2+\zeta,-1+2\zeta;3+\zeta;-1\right)\nonumber\\
&+&
\frac{\Gamma(\zeta-3)\Gamma(2-2\zeta)}{\Gamma(-1-\zeta)}
\frac{(2-\zeta)(2\zeta-3)}{\zeta}
-\frac{1}{1-\zeta}F\left(-1+\zeta,-1+2\zeta;\zeta;-1\right)\nonumber\\
&+&\frac{(1-2\zeta)}{2-\zeta}
F\left(-2+\zeta,-1+2\zeta;-1+\zeta;-1\right)
-\frac{1}{3-\zeta} F\left(-3+\zeta,-1+2\zeta;-2+\zeta;-1\right)
\nonumber\\
&+&\frac{x^\zeta}{\zeta} \frac{2(1-2\zeta)(2\zeta^3-5\zeta^2+2)}{(\zeta-2)}
+{\cal O}\left(x^{\zeta+2}\right).\label{Ffullsim}
\end{eqnarray}
Thus for small $x=p/\Lambda$, the ${\cal F}$ can be written as
\begin{eqnarray}
{\cal F}(\zeta,x)&\simeq& {\cal F}(\zeta)+{\cal O}\left(x^\zeta\right),
\label{Fzeta}
\end{eqnarray}
where we write ${\cal F}(\zeta)={\cal F}(\zeta,0)$.
For values of $\zeta$ with $0<\zeta <1/2$, 
the self energy has the scaling form
\begin{eqnarray}
\Sigma_A(p)\simeq
-\lambda(\zeta)\left(\frac{p}{\Lambda}\right)^{-\zeta},
\label{scalfsigA}
\end{eqnarray}
where
\begin{eqnarray}
\lambda(\zeta)\equiv \frac{
{\cal F}(\zeta)}{8 {\cal B}(\zeta) (3-2\zeta)(1-2\zeta)}.
\label{lameqdef}
\end{eqnarray}
Using the scaling form Eq.~(\ref{scalfsigA}) and the Ansatz
(\ref{wavefieansatz}) in Eq.~(\ref{Aeq}), we get the self-consistency relation
\begin{eqnarray}
\left(\frac{p}{\Lambda}\right)^{-\zeta}&=&1-N\Sigma_A(p)=
N\lambda(\zeta)
\left(\frac{p}{\Lambda}\right)^{-\zeta}
+{\cal O}(1).
\end{eqnarray}
This gives the eigenvalue equation (for $0<\zeta<1/2$)
\begin{equation}
\lambda(\zeta)=1/N\label{lameq}.
\end{equation}
This equation fixes the dependence of the
anomalous dimension $\zeta$ on $N$.  

In the limit $\zeta\to 0$,
the function ${\cal F}$ given by Eq.~(\ref{Ffullsim}) 
reduces to
\begin{eqnarray}
{\cal F}(0,x)\approx -2\ln x. 
\label{Fzeta0}
\end{eqnarray}
At $\zeta=0$, using ${\cal B}(0)=\pi^2/8$ and
Eq.~(\ref{Fzeta0}), 
we find that 
$\Sigma_A(p)\approx
2/(3\pi^2)\ln (p/\Lambda)$ which is consistent with the one-loop or leading
$1/N$ correction to $\Sigma_A$ in the Gross-Neveu model \cite{hakoko93}.
At $\zeta=1/2$ the function ${\cal F}(\zeta)$ has a first order root
and we can extract the residue by taking the limit $\zeta\to 1/2$ 
in Eq.~(\ref{calFdef}) (at $x=0$)
\begin{eqnarray}
\lim_{\zeta\to 1/2}\frac{{\cal F}(\zeta)}{1-2\zeta}
&=&
\int\limits_0^1dt\,\left[t^{-1/2}(1+t^2)\ln\frac{1+t}{1-t}-2t^{1/2}\right]
\nonumber\\
&+&
\int\limits_0^1dt\,
\left[t^{-7/2}(1+t^2)\ln\frac{1+t}{1-t}-2t^{-5/2}\right]\nonumber\\
&=&\frac{12\pi}{5}.\label{Fnearzhalf}
\end{eqnarray}
Thus $F(\zeta)\approx 12\pi(1-2\zeta)/5$ for $\zeta\uparrow 1/2$.
Near $\zeta=1/2$, we obtain by using Eq.~(\ref{Bnearzhalf}), 
Eq.~(\ref{lameqdef}), and Eq.~(\ref{Fnearzhalf}) that
\begin{eqnarray}
\lambda(\zeta)\approx \frac{2\pi(1-2\zeta)}{5}=\frac{1}{N}
\quad \Longrightarrow\quad
\zeta\approx \frac{1}{2}-\frac{5}{4\pi N}.\label{zetaapprox}
\end{eqnarray}

\begin{figure}
\rotatebox{-90}{
\resizebox*{0.5\columnwidth}{!}{\includegraphics{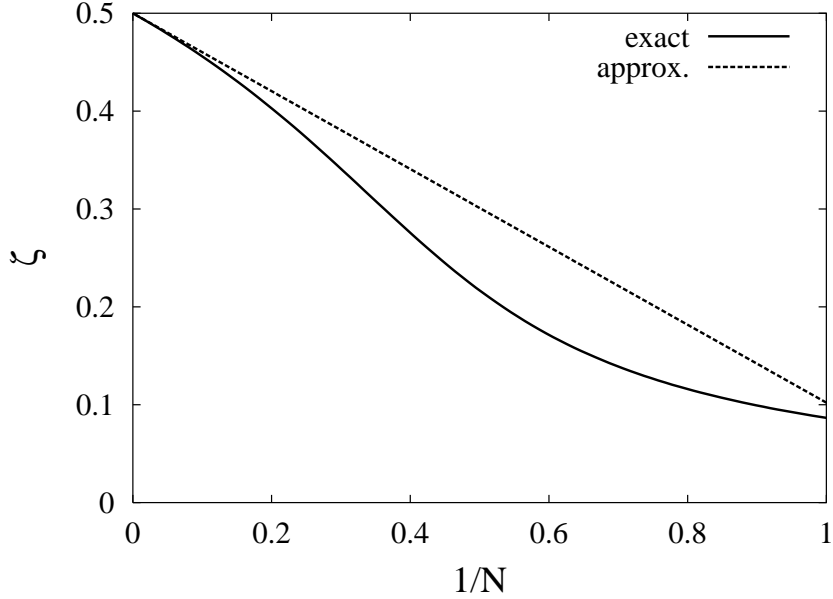}}}
\caption{Exact solution of Eq.~(\ref{lameq}) compared with
$\zeta=1/2-5/(4\pi N)$; 
the exponent $\zeta$ is plotted versus $1/N$.}
\label{fig_zetavsinvN}
\end{figure}

The exact solution Eq.~(\ref{lameq}) and the above approximate solution
are depicted in Fig.~\ref{fig_zetavsinvN}. 
Clearly Fig.~\ref{fig_zetavsinvN} shows that if
$N\rightarrow \infty$ then $\zeta\rightarrow 1/2$.  
This behavior is a result of the fact that the functions 
$1/{\cal B}(\zeta)$ as defined in
Sec.~\ref{sec_scalbeh} and the function ${\cal F}(\zeta)$ defined 
in Eq.~(\ref{Fzeta}) have a first order root at $\zeta=1/2$.
There, the eigenvalue equation is of the form given by 
Eq.~(\ref{zetaapprox}).

Finally we mention that the leading scaling behavior is independent of
the choice of momenta flowing in the self energy diagram $\Sigma_A$
given by Eq.~(\ref{sigAdef}); an alternative choice, {\em e.g.}, $k+p$
for the fermion propagator and $k$ for $\Delta_\sigma$ leads to the
same scaling form (\ref{scalfsigA}) and eigenvalue equation
(\ref{lameq}).
\section{The gap equation and the critical exponents}\label{sec_gapeq}
In the previous section the scaling laws for $A(p)$ and
$\Delta_\sigma(p)$ at the critical coupling $g=g_c$ were obtained.
This yielded the anomalous dimension $\eta=1-2\zeta$ with $\zeta$
given by Eq.~(\ref{lameq}).  In order to obtain the other critical
exponents the set of SD equations should be studied either in the
subcritical ($g\uparrow g_c$) or supercritical ($g\downarrow g_c$)
regime. In these regimes the analysis of the SD equations depicted in
Fig.~\ref{fig_trunc} is more complicated, due to the fact that a
nonzero infrared mass scale (the inverse correlation length) enters
the problem.  For instance in the subcritical regime the inverse
correlation length or $m_\sigma$ is given by the complex pole $p^2\sim
m_\sigma^2 \exp(-i\theta)$ in $\Delta_\sigma(p)$ \cite{guhare01}. The
critical exponent $\nu$ can then be derived from the scaling law
$m_\sigma\sim (g_c-g)^\nu$.  If we consider Eqs.~(\ref{Delsprop}) and
(\ref{Pisigexpr}) in the subcritical regime, it follows that
$\nu=1/(1+2\zeta)$. Consequently the anomalous dimension $\gamma=1$ as
can be determined from the limit $\Delta_\sigma(p\rightarrow 0)\sim
1/|g_c-g|\sim \chi$ with $\chi$ the usual susceptibility
\cite{hakoko93,re9900}.

In the subcritical regime the $\sigma$ boson propagator will be of the 
renormalizable form:
\begin{eqnarray}
\Delta_\sigma(p)\simeq-\frac{\pi^2}{2\Lambda {\cal B}(\zeta)}
\left(\frac{\Lambda}{p}\right)^{1+2\zeta}\frac{1}{\left[1
+\left(m_\sigma^2/p^2\right)^{(1+2\zeta)/2}\right]}, \label{bospropsubcrit}
\end{eqnarray}
where $m_\sigma$ is the generic notation for inverse correlation length
($m_\sigma\propto m_{dyn}\sim 1/\xi$).
If this expression (\ref{bospropsubcrit}) is inserted in Eq.~(\ref{sigAdef})
a careful analysis will show that the fermion wave function $A$ 
behaves as follows in the subcritical regime,
\begin{equation}
A(p)\simeq \left(\frac{\max(p,c_{-} m_\sigma)}{\Lambda}\right)^{-\zeta},
\label{Asubcrit}
\end{equation}
where $c_{-}$ is a nonuniversal constant. 
This constant $c_{-}$ should be determined from the nonlinear SD equations 
(\ref{Delsprop})--(\ref{sigAdef}),
but for the scaling arguments its precise value is irrelevant.
In the supercritical regime the wave function will be of the same form as 
Eq.~(\ref{Asubcrit}), but with a different 
constant of proportionality, {\em e.g.}, $c_{+}$ instead of $c_{-}$.

In rest of this section the supercritical regime ($g>g_c$)
is considered and we
study the dynamical symmetry breaking SU($N$)$\,\rightarrow\,$U(1).
As was explained in Sec.~\ref{sec_largeN}, in the leading large $N$
approximation only the tad-pole graph is included in the gap equation
for $M$.  This gap equation with bare mass $h$ reads (in Euclidean
formulation)
\begin{eqnarray}
M=h+4G\int_E\frac{d^3p}{(2\pi)^3}\,\frac{M}{A^2(p)p^2+M^2}.
\label{gapeq}
\end{eqnarray}
For the wave function $A$ the solution 
which is valid in the broken phase should be substituted. 
However, since the scaling laws for the order parameter and dynamical mass 
are mainly determined by the ultraviolet behavior 
$m_\sigma\ll p \leq \Lambda$ of the theory, we can safely 
use Eq.~(\ref{wavefieansatz}).
Since the mass $M$ in Eq.~(\ref{gapeq})
is independent of momentum, the above integral 
can be performed.  Using Eq.~(\ref{wavefieansatz}) for the fermion
wave function $A$, we obtain
\begin{eqnarray}
1\simeq\frac{h}{M}+g \left(\frac{\Lambda^2}{3M^2}\right){\,_2F_1}
\left(1,\frac{3}{2-2\zeta};1+\frac{3}{2-2\zeta};-\frac{\Lambda^2}{M^2}\right).
\end{eqnarray}
The argument of the hypergeometric function approaches $-\infty$ as
the critical curve is approached ($g\rightarrow g_c$, $M\rightarrow
0$) at $h=0$.  Therefore, close to criticality, the gap equation can
be approximated as follows:
\begin{eqnarray}
-\left(\frac{g-g_c}{g_c}\right)=\frac{h}{M}
+\frac{g}{3}\Gamma\left(1+b\right)
\Gamma\left(1-b\right)
\left(\frac{M}{\Lambda}\right)^{2b-2},\label{gap3}
\end{eqnarray}
where $b=3/(2-2\zeta)$ and $g_c=1+2\zeta$.

Note that within our approximation $M=\langle \sigma_3\rangle$.
Therefore, treating $M$ as the order parameter, the gap equation
(\ref{gap3}) is the quantum mechanical equivalent of the equation of
state known in statistical mechanics and in analogy various critical
exponents can be derived from it (see
Refs.~\cite{kohakoda90balomi90,hakoko93}).  For instance, the critical
exponent $\delta$ follows from the scaling law $h\sim M^\delta$ at
$g=g_c$;
\begin{eqnarray}
h\propto M^{2b-2+1}\quad \Longrightarrow \quad \delta=2b-1
=\frac{2+\zeta}{1-\zeta}
\end{eqnarray}
and the critical exponent $\beta$ 
follows from $M\sim (g-g_c)^\beta$ at $h=0$;
\begin{equation}
g-g_c\sim M^{2b-2}\quad \Longrightarrow \quad
\beta=\frac{1}{2b-2}=\frac{1-\zeta}{1+2\zeta}.
\end{equation}
In a similar manner the critical exponent $\gamma$ can be computed
from $\chi=\partial M/\partial h|_{h=0}\sim (g-g_c)^{-\gamma}$;
\begin{equation}
\frac{\partial M}{\partial h}\biggr|_{h=0}\sim M^{2-2b}\sim (g-g_c)^{-1}
\quad\Longrightarrow \quad 
\gamma=1.
\end{equation}
The other critical exponents can be derived from the effective
potential and the scaling form for the composite boson propagator
Eq.~(\ref{bospropcrit}).  Summarizing, the critical coupling $g_c$ and
the critical exponents are expressed in terms of the exponent $\zeta$
of the fermion wave function $A$ which is determined by
Eq.~(\ref{lameq});
\begin{eqnarray}
&&g_c=1+2\zeta,\qquad 
\beta=(1-\zeta)/(1+2\zeta),\qquad\delta=(2+\zeta)/(1-\zeta),\nonumber\\
&&\nu=1/(1+2\zeta),
\qquad \gamma=1,\qquad \eta=1-2\zeta. \label{univclass}
\end{eqnarray}
These exponents
satisfy the three dimensional hyperscaling equations
$\gamma=\beta(\delta-1)$, $\gamma=\nu(2-\eta)$, and $3\nu=2\beta+\gamma$.  
The critical
exponent $\nu$ follows from the scaling law for the dynamical mass $m_{dyn}$
(inverse correlation length)
describing the physical fermion mass in the broken phase close to
criticality; $m_{dyn}\sim M^{1/(1-\zeta)}$ (see Sec.~\ref{sec_model}).

In the limit $N\rightarrow \infty$,
keeping only terms to order $1/N$ 
by using the expression (\ref{zetaapprox}) for $\zeta$, 
the critical exponents are
\begin{eqnarray}
\beta=\frac{1}{4}+\frac{15}{16\pi N},\quad \delta=5-\frac{15}{\pi N},
\quad \nu=\frac{1}{2}+\frac{5}{8\pi N},\quad 
\gamma=1, \quad \eta=\frac{5}{2\pi N}.\label{sumofexp}
\end{eqnarray}
These exponents clearly differ from the next-to-leading large $N$ exponents of the
Gross-Neveu model \cite{hakoko93};
\begin{eqnarray}
\beta=1,\quad \delta=2+\frac{8}{\pi^2N}, 
\quad \nu=1+\frac{8}{3\pi^2N},\quad \gamma=1+\frac{8}{N\pi^2}, \quad 
\eta=1-\frac{16}{3\pi^2 N}.
\end{eqnarray}
Note, for instance, the fractions of $1/(\pi N)$ appearing in the
exponents (\ref{sumofexp}), whereas the Gross-Neveu exponents contain
fractions of $1/(\pi^2 N)$.

From Eq.~(\ref{wavefieansatz}) and (\ref{bospropcrit}), it follows
that the anomalous dimension $\eta_\psi$ of the fermion field is
$\eta_\psi=-\zeta/2\approx
-1/4+5/(8\pi N)$ and that the anomalous dimension $\eta_\sigma$ of
the composite boson field is $\eta_\sigma=\zeta\approx 1/2-5/(4\pi N)$.  Since the
$\beta$-function of $g$ vanishes at critical coupling $g=g_c$ ({\em
i.e.}, $\beta(g_c)=0$), the Yukawa vertex should satisfy the following
Callan-Symanzik equation at $g=g_c$:
\begin{equation}
\left(\eta_\sigma+2\eta_\psi
+\Lambda\frac{\partial}{\partial \Lambda}\right)\Gamma^A(k+p,k)\approx 0.
\end{equation}
Clearly, the bare Yukawa vertex approximation 
$\Gamma^A(p+q,p)=\tau^A$
is consistent with this scaling constraint, since 
the scaling dimension of the bare vertex is zero in agreement with the
equality $\eta_\sigma=-2\eta_\psi$.
\section{Discussion}\label{sec_dis}
\begin{figure}
\resizebox*{0.85\columnwidth}{!}{\includegraphics{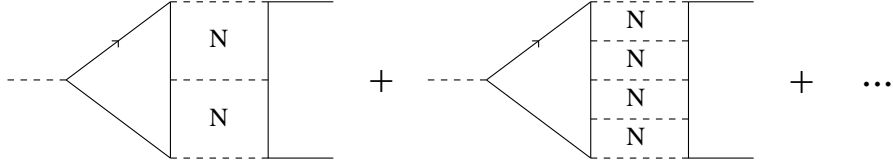}}
\caption{Next-to-leading order $1/N$ corrections to the Yukawa vertex}
\label{fig_next2leading}
\end{figure}

In Eq.~(\ref{sumofexp}), we have summarized the critical exponents
obtained within our bare vertex approximation to order $1/N$.
Together with the anomalous dimensions $\zeta=1/2-5/(4\pi N)$ and
$\eta=1-2\zeta$ for respectively the fermion and $\sigma$ boson
propagators for the model presented, this is the main result of this
paper.  However, the truncation scheme depicted in
Fig.~\ref{fig_trunc} describes merely the leading behavior in the
$1/N$ expansion.  Interestingly, the solutions of the truncated system
depicted in Fig.~\ref{fig_trunc} satisfy the hyperscaling equations
for arbitrary but finite $N$.  However, we cannot say with certainty
that the values obtained in Eq.~(\ref{sumofexp}) are the correct
critical exponents to order $1/N$, without a thorough investigation of
the ${\cal O}(1/N)$ diagrams for the Yukawa vertex.  Examples of such
$1/N$ Yukawa vertex corrections are depicted in
Fig.~\ref{fig_next2leading}.
The factors of $N$ depicted in this figure represent index-loops which
can be straightforwardly determined, using 't Hoofts double line
representation (as explained in the Appendix).

Suppressing factors of $1/N$ appear in the following way.  From
section~\ref{sec_scalbeh}, using Eqs.~(\ref{bospropcrit}) and
(\ref{Bnearzhalf}), we find that the $\sigma$ boson propagator at the
critical coupling has the approximate form
\begin{eqnarray}
\Delta_\sigma(p)\simeq-\frac{4\pi^2 (1-2\zeta)}{3\Lambda }
\left(\frac{\Lambda}{p}\right)^{1+2\zeta}. \label{bospropcrit2}
\end{eqnarray}
The eigenvalue equation (\ref{zetaapprox}) identifies a factor of
$1/N$ with $1-2\zeta$ ($1-2\zeta\propto 1/N$).  Hence, each internal
full $\sigma$ boson line in a Feynman diagram carries a suppressing
factor $1/N$. This illustrates that the diagrams in
Fig.~\ref{fig_next2leading} are indeed of order $1/N$.  Now the factor
$1-2\zeta$ or $1/N$ is related to the Yukawa coupling $g_Y$ mentioned
in the Appendix.  Since a four-fermion scattering amplitude,
exchanging a virtual $\sigma$ boson would correspond to a factor
$g_Y^2$, we can associate the square of the Yukawa coupling with
$1/N$, $g_Y^2\sim 1/N$, so that the factors $g_Y^2 N$ are indeed of
order one, in retrospect.  Using the above observation that each
internal $\sigma$ boson propagator corresponds to a factor $1/N$ and
each index-loop to a factor $N$, it is straightforward to verify that
the vertex corrections depicted in Fig.~\ref{fig_next2leading} are
order $1/N$.  Moreover, diagrams containing fermion loops with an odd
number of out-going $\sigma$ lines are zero in the subcritical regime,
due to restrictions imposed by the symmetry of the model.

If $N=\infty$, we should get the exponents within the accuracy of the
approximation.  It turned out, however, that at $N=\infty$,
$\zeta=1/2$ and no pure power-law expression exists for the $\sigma$
boson propagator (see Eq.~(\ref{Pilog})). This logarithmic deviation
from the power-law behavior causes the violation of hyperscaling and
consequently spoils the renormalizability of the model.  Such a
violation could be a sign that the $1/N$ expansion is at the most an
asymptotic expansion, since continuation to negative values of $1/N<0$
are physically meaningless.  But at the moment, this is a mere
hypothesis.

Nevertheless, when we take finite but large $N$ then the model does
have power-law renormalizability and respects hyperscaling.  The main
question remains whether the exponents in Eq.~(\ref{sumofexp}) are
correct to order $1/N$. In order to settle this question adequately
one needs to investigate the $1/N$ corrections to the Yukawa vertex.
We can think of three scenarios.  1. The next-to-leading $1/N$
corrections as depicted in Fig.~\ref{fig_next2leading} are finite and
will contribute only to the ``conformal invariant'' or scale invariant
part of the Yukawa vertex and do not alter its scaling dimension.
This so-called scale invariant part of the Yukawa vertex depends only
on the ratios of the in- and out-going momenta.  2. Each of the
diagrams depicted in Fig.~\ref{fig_next2leading} is logarithmically
divergent and the infinite sum of such diagrams give rise to a new
scaling dimension of order $1/N$ for the Yukawa vertex.  3. The $1/N$
expansion breaks down due to the appearance of kinematic factors such
as $1/(1-2\zeta)$ terms, which give rise to additional factors of $N$
and render the whole idea of topological $1/N$ expansion meaningless.
In case that scenario 1 would turn out to be correct, then the set of
critical exponents (\ref{sumofexp}) will be correct to order $1/N$.
Scenario 2 leads to an anomalous dimension for the Yukawa vertex of
order $1/N$ and the expressions for the critical exponents and
anomalous dimension are only valid in the leading order and up to
logarithmic accuracy.  In case that scenario 3 unfolds, our analytical
approach has failed, since then we lost track of the perturbative
parameter $1/N$ and no standard techniques, except perhaps lattice
simulations, are adequate to tackle the critical behavior of the
presented model.  By the way, each of three scenarios can be
reconciled with the full conformal symmetry at the critical point of
four-fermion models as advocated in Ref.~\cite{chmase93}.

\section{Summary and conclusion}\label{sec_concl}
In this paper we have studied various aspects of dynamical symmetry
breaking SU($N$)$\,\rightarrow\,$U(1) in a specific SU($N$)
four-fermion model in the large $N$ limit in 2+1 dimensions, by making
use of 't Hoofts topological $1/N$ expansion.  We advocated the
applicability of 't Hoofts topological $1/N$ expansion to the
presented model.  Within the adopted $1/N$ expansion, the Yukawa
vertex was approximated by its bare form.  The main result of this
paper is the determination of the exponent $\zeta$ or anomalous
dimension of the fermion wave function $A(p)$ as a function of $N$ in
the planar approximation at the critical point $g=g_c$.  The
functional relationship between $\zeta$ and $N$ given in
Eq.~(\ref{lameq}) and Fig.~\ref{fig_zetavsinvN}.  The approximate
relationship was found to be $\zeta\simeq 1/2-5/(4\pi N)$.
Subsequently the critical coupling $g_c$ and various critical
exponents are given in terms of the exponent $\zeta$.  In the large
$N$ limit, the critical exponents for this SU($N$) invariant
four-fermion model differ considerably from the critical exponents of
the Gross-Neveu model.  The presence of $N^2-1$ light composite bosons
turned out to drastically changes the scaling behavior of the fermion
propagator in the vicinity of the critical point, giving rise to a new
three dimensional universality class, see Eq.~(\ref{univclass}).
Similar to the 2+1 dimensional Gross-Neveu model, the presented model
is nonperturbatively renormalizable for large but finite $N$.  The
obtained anomalous dimensions of the fermion and $\sigma$ boson
propagators are self-consistent with the bare Yukawa vertex
approximation.

For the future, it could be interesting to extend the presented
approach to arbitrary space-time dimension $d$ in order to investigate
the perseverance of triviality at the upper critical dimension
$d=3+1$.
\acknowledgments{The author wishes to thank V.P.~Gusynin for useful
discussions and suggestions.}
\appendix*
\section{The topological \boldmath{$1/N$} expansion}\label{ap_1overN}
In this appendix, we illustrate the applicability of 't Hoofts
topological $1/N$ expansion to our model Lagrangian (\ref{sunffm}).
As is explicitly mentioned in Ref.~\cite{tho74}, the topological
expansion is also valid for models, with a global U($N$) symmetry,
having scalar fields with two U($N$) indices ({\em i.e.}, the adjoint
representation). Therefore, for large $N$, we generalize the SU($N$)
symmetry to U($N$). Subsequently, we perform a Hubbard-Stratonovich
transformation, after which the Lagrangian (\ref{sunffm}) is expressed
in terms of Hermitian scalar fields $\sigma_{\alpha\beta}
=\sigma^\ast_{\beta\alpha}$,
\begin{eqnarray}
{\cal L}&=&
\bar\psi_\alpha i\hat \partial\psi_\alpha
-\bar\psi_\alpha\psi_\beta\sigma_{\alpha\beta}
-\frac{1}{2G}\sigma_{\alpha\beta}\sigma_{\beta\alpha},\label{HSsunffm2}
\end{eqnarray}
with the flavor or ``color'' indices $\alpha$, $\beta$ running from
$1$ to $N$. Contrary to the QCD and U($N$) gauge theories,
``physical'' observables are not restricted to be ``color'' blind.  In
other words, external fields or sources are allowed to carry U($N$)
flavor indices.  The Hermitian scalar field are linearly related to
the $N^2$ real scalar fields $\sigma^A$ (for U($N$)) of
Eq.~(\ref{HSsunffm}),
\begin{eqnarray}
\sigma_{\alpha\beta}=\sum_{A=0}^{N^2-1}\sigma^A \tau^A_{\alpha\beta},
\qquad \sigma^A=\tau^A_{\alpha\beta}\sigma_{\beta\alpha},
\label{hersc_realsc}
\end{eqnarray}
with the U(1) generator
$\tau^0_{\alpha\beta}=\delta_{\alpha\beta}/\sqrt{N}$.

Now the scalar fields $\sigma_{\alpha\beta}$ are analogous to the
gluon gauge fields of Ref.~\cite{tho74}, carrying two U($N$) indices,
and consequently the topological $1/N$ expansion can be applied.
Moreover, due to the absence cubic and quartic self interactions for
the auxiliary scalar fields, the ``flavor'' Feynman rules are
considerably simpler than the corresponding rules for the U($N$) gauge
field theories.
\begin{figure}
\resizebox*{0.45\columnwidth}{!}{\includegraphics{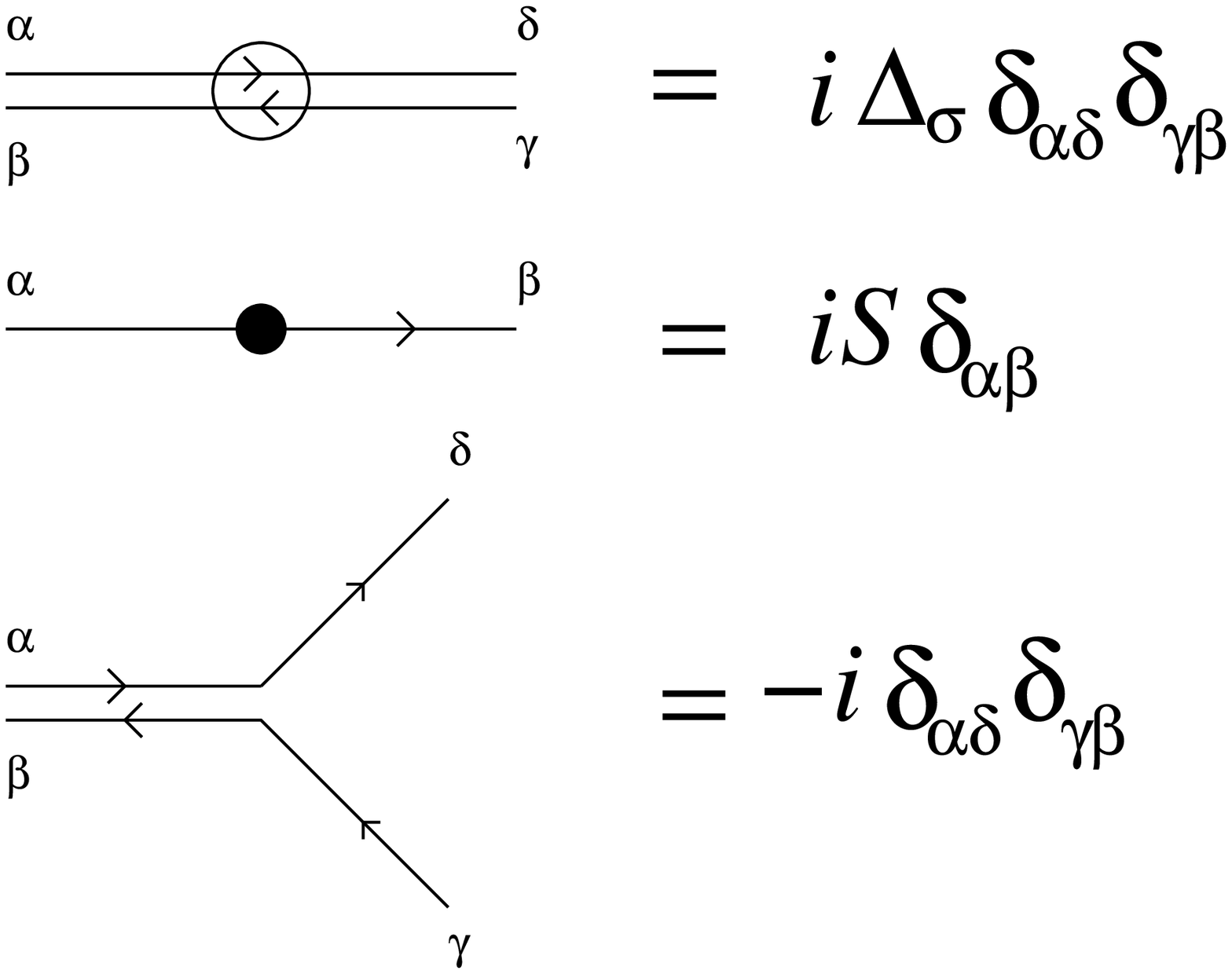}}
\caption{The double index Feynman rules for the symmetric phase, with
$\Delta_\sigma^{-1}(p)=-1/G+\Pi_\sigma(p)$ and $S^{-1}(p)=\hat
p-\Sigma(p)$, where $\Sigma$ is fermion self-energy.  }
\label{fig_flavorfrul}
\end{figure}
These flavor Feynman rules for the U($N$) symmetric phase are depicted
in Fig.~\ref{fig_flavorfrul}, using the double line representation of
't Hooft, see also Ref.~\cite{fo00afo00b}.  Each index loop in a
Feynman diagram gives rise to factor $N$.  In
Fig.~\ref{fig_flavorfrul}, the propagators are fully dressed.  This is
necessarily so, because of the fact that in a critical theory, with
relevant four-fermion interactions close to or at a critical point,
the vacuum polarization $\Pi_\sigma$ is of a comparable magnitude as
the bare mass term $1/G$. The fermion self-energy $\Sigma$ is of the
order of the canonical or free form $\hat p$ or larger (this will be
shown in Sec.~\ref{sec_scalbeh}).  At the critical point the bare mass
term $1/G$ of the $\sigma$ boson propagator $\Delta_\sigma(p)$ is
canceled by the momentum independent $\sigma$ boson vacuum
polarization ($\Pi_\sigma$) corrections and the remaining momentum
dependent terms give rise to the anomalous scaling law for the scalar
propagator.  For the fermion propagator a similar argument holds, see
also Sec.~\ref{sec_scalbeh}.

For $d=4$ the leading $1/N$ expansion for U($N$) gauge theories
consists of complicated planar networks of gluon exchanges.  This is
mainly due to the local nature of the symmetry, which introduces cubic
and quartic gluon self-couplings and gauge-fixing interactions, whose
particular structure depends on the space-time dimension.  For $d=2$,
't Hooft showed that the set of planar diagrams can be reduced to
self-energy and ladder diagrams.  As a consequence, the SDEs form a
closed set and the $d=2$ gauge theory is solvable in the leading $1/N$
expansion.

Contrary to the local U($N$) symmetries, the topological $1/N$
expansion is not space-time dependent for the global U($N$) symmetries
of our interest, merely because of the absence of gauge fixing
constraints or Faddeev-Popov ghosts.  The space-time dimension $d$
does not play a role in the selection rules for the planarity and thus
the large $N$ behavior of Feynman diagrams for global U($N$).  For
internal $\sigma$ boson exchanges is does not matter whether we use
the real representation or the Hermitian representation for the
composite fields to leading order in $1/N$, since
\begin{eqnarray}
\langle \sigma_{\alpha\beta}(p)\sigma_{\gamma\delta}(-p)\rangle_C
&=&\sum_A \sum_B \tau^A_{\alpha\beta}\tau^B_{\gamma\delta} \langle
\sigma^A(p)\sigma^B(-p)\rangle_C =
\sum_A\tau^A_{\alpha\beta}\tau^A_{\gamma\delta}
i\Delta^{A}_\sigma(p)\nonumber \\
&=&i\Delta_\sigma(p)\delta_{\alpha\delta}\delta_{\gamma\beta}
+\frac{1}{N}\left[i\Delta^0_\sigma(p)-i\Delta_\sigma(p)\right]
\delta_{\alpha\beta}\delta_{\gamma\delta},\label{delshermreal}
\end{eqnarray}
where the $N^2-1$ propagators $\Delta^A_\sigma$ are degenerate,
$\Delta^A_\sigma=\Delta_\sigma$, and $\Delta^0_\sigma$ is the
propagator of the U(1) field $\sigma^0$.  In general
$\Delta^0_\sigma\not =\Delta_\sigma$, thus from
Eq.~(\ref{delshermreal}) it follows that the double line
representation of the $\sigma$ boson propagator is violated at the
most by a $1/N$ suppressed term.

\begin{figure}
\resizebox*{0.65\columnwidth}{!}{\includegraphics{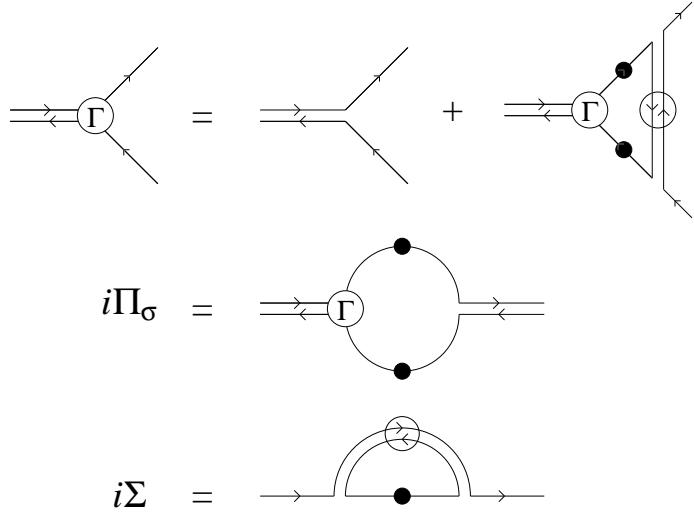}}
\caption{The planar approximation for the Yukawa vertex
$\Gamma_\sigma$, the $\sigma$ boson vacuum polarization $\Pi_\sigma$
and the fermion self energy $\Sigma$.}
\label{fig_sdeplanar}
\end{figure}

\begin{figure}
\resizebox*{0.65\columnwidth}{!}{\includegraphics{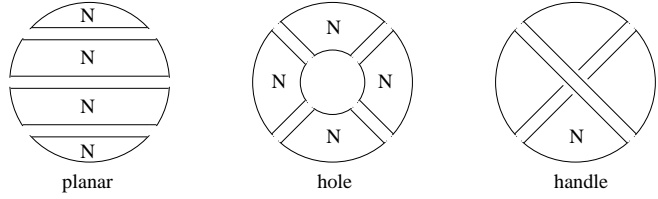}}
\caption{The planar diagram corresponds to a factor $N (g_Y^2 N)^3$.
The hole diagram corresponds to a factor $(g_Y^2 N)^4$, and the handle
diagram to a factor $1/N (g_Y^2 N)^2$.}
\label{fig_planholhan}
\end{figure}

The planar expansion for the present 2+1 dimensional model is
equivalent to the planar expansion for the $d=2$ U($N$) gauge theory
or 't Hooft model, since both models have a single interaction vertex,
see Fig.~\ref{fig_sdeplanar}.
This planarity can be verified order by order in the diagrammatic
expansion of full $\sigma$ propagators $\Delta_\sigma$.  We assume
that with each internal full $\sigma$ boson exchange in a Feynman
diagram an ``effective'' Yukawa coupling, $g_Y$ is associated. We
assume that combination $g_Y^2 N$ is fixed and of the order of one
\cite{re9900}, so that all diagrams with powers of $g^2_Y N$ are of
the same order and need to be summed nonperturbatively.  The
assumption that $g_Y^2 N$ is fixed for $N\rightarrow \infty$ can only
be validated in retrospect, since $g_Y$ is not a free parameter like
the gauge coupling $g$ in QCD; $g_Y$ is self-consistently determined
from the planar SDEs and therefore depends on $N$ and $G$.  For the
$\sigma$ boson vacuum polarization $\Pi_\sigma$, each planar diagram
with $n$ internal exchanges of $\sigma$ bosons is associated with a
factor $N^ng_Y^{2n}$.  For the fermion self-energy $\Sigma$, each
planar diagram with $n$ $\sigma$ exchanges also corresponds to a
factor $N^n g_Y^{2n}$. Examples of vacuum diagrams with different
topology (planar, hole, handle) are depicted in
Fig.~\ref{fig_planholhan}.

For the fermion self energy, the planar expansion is generated by the
rainbow approximation.  Moreover for the present model, a certain
class of planar diagrams can be neglected.  This class comprises all
planar diagrams, which can be identified as vertex corrections.  This
can be understood in the following way.  As depicted in
Fig.~\ref{fig_sdeplanar}, a full Yukawa vertex can either connect the
left upper in-going line in $\Pi_\sigma$ with a right upper outgoing
line, or connect the upper left line with the lower left line. Whereas
the bare vertex corresponds to the former, the ladder vertex
corrections correspond to the latter.  Since the generator $\tau^0$ is
the only generator having a nonzero trace, it is straightforward to
check that diagrams connecting upper left lines with lower left lines
give only nonzero contributions to the propagator $\Delta^0_\sigma$ of
the U(1) field, thereby causing the inequality $\Delta^0_\sigma\not =
\Delta_\sigma$. For instance, in the real representation, the vacuum
polarization $\Pi^A_\sigma$ of the propagator $\Delta^A_\sigma$ gets a
two-loop vertex correction which is proportional to $\sum_B {\rm Tr\,}
\tau^A\tau^B\tau^A\tau^B$.  This trace vanishes, except for $A=0$.
Using Eq.~(\ref{delshermreal}) in the fermion self energy, it follows
that the contribution of $\Delta^0_\sigma$ is suppressed by a factor
$1/N^2$ with respect to the degenerate SU($N$) propagator
$\Delta_\sigma$.  Therefore, we assume that all relevant planar
diagrams, {\em i.e.}, diagrams contributing to $\Delta_\sigma$, are
given by the bare vertex approximation depicted in
Fig.~\ref{fig_barevertex}.  A similar planar diagram approach for the
SU($N$) extrapolation of the Hubbard model was adopted by
Foerster~\cite{fo00afo00b}.
\end{document}